\RequirePackage{amsmath}
\documentclass[journal]{IEEEtran}
\ifCLASSINFOpdf
\else
\fi
\usepackage[british]{babel}
\usepackage{comment}
\usepackage{color}
\usepackage{amsfonts}
\usepackage{graphicx}

\newtheorem{theorem}{Theorem}[section]
\newtheorem{definition}[theorem]{Definition}
\newtheorem{lemma}[theorem]{Lemma}
\newtheorem{problem}[theorem]{Problem}
\newtheorem{assumption}[theorem]{Assumption}
\newtheorem{remark}[theorem]{Remark}

\newtheorem{corollary}[theorem]{Corollary}

\begin{document}

\title{Traffic Models of Periodic Event-Triggered Control Systems}

\author{Anqi~Fu,~and~Manuel~Mazo,~Jr.
\thanks{The authors are with the Delft Center for Systems and Control, Delft University of Technology, Delft 2628 CD, The Netherlands e-mail: {\tt\small {A.Fu-1, M.Mazo}@tudelft.nl}.}
\thanks{This work is partly funded by China Scholarship Council (CSC).}

}
\maketitle              

\begin{abstract}
Periodic event-triggered control (PETC) \cite{heemels2013periodic} is a version of event-triggered control (ETC) that only requires to measure the plant output periodically instead of continuously. In this work, we present a construction of timing models for these PETC implementations to capture the dynamics of the traffic they generate. In the construction, we employ a two-step approach. We first partition the state space into a finite number of regions. Then in each region, the event-triggering behavior is analyzed with the help of LMIs. The state transitions among different regions result from computing the reachable state set starting from each region within the computed event time intervals.
\end{abstract}

\ifCLASSOPTIONpeerreview
\begin{center} \bfseries EDICS Category: 3-BBND \end{center}
\fi
\IEEEpeerreviewmaketitle

\section{Introduction}

Wireless networked control systems (WNCS) are control systems that employ wireless networks as feedback channels. In such systems, the physically distributed components are co-located with their own wireless nodes and communicate via a wireless network. These components can be designed with great mobility once the nodes are supported by batteries. Besides, each component can be established and updated easily. Therefore, WNCS have great adaptability on obtaining different control objectives and have been attracting much attention. However, there are two major issues that must be considered while designing such a system: limited bandwidth and energy supply.

Most often, control tasks are designed to be executed periodically. This periodic strategy, also named time-triggered control (TTC), does not regard the system's current state and thus may waste bandwidth and energy. Alternatively, event-triggered control (ETC) strategies are proposed to reduce bandwidth occupation, see e.g. \cite{donkers2012output}, \cite{MazoJr2014}, \cite{mazo2011decentralizedtac}, \cite{tabuada2007event}, \cite{wang2011event}, \cite{wang2011eventautomatica}, and references therein. In ETC, the control tasks only execute when necessary, e.g. when some pre-designed performance indicator is about to be violated. Thus the system is tightfisted in communication. However, to validate the pre-designed event-triggering conditions, sensors are required to sample the plant output continuously. This continuous monitoring can consume large amounts of energy. To reduce this energy consumption, naturally one may want to replace the continuously sampling by a discrete time sampling.

When applying discrete time sampling, to compensate the delay caused by the discretization, one can either design a stricter event-triggering condition based on the system dynamics, as e.g. \cite{mazo2015decentralized}; or modify the Lyapunov function, as e.g. \cite{heemels2013periodic}. In \cite{heemels2013periodic}, Heemels et. al. present a periodic event-triggered control (PETC) mechanism. In a PETC implementation, the sensors are only required to measure the plant output and validate the event conditions periodically. Only when some pre-designed conditions are satisfied, fresh measurements are employed to recompute the controller output. Therefore, PETC enjoys the benefits of both cautious communication and discrete time measurement. Compared to \cite{mazo2015decentralized}, the event conditions can be less conservative to further reduce communications. Thus the energy consumed and bandwidth occupied are reduced. Furthermore, the transmissions of the control input from the controller to the plant are also included in the PETC mechanism.

To further reduce the resource consumption and to fully extract the potential gains from ETC, one can also consider scheduling approaches. By efficiently scheduling listening times on wireless communications and medium access time in general, the energy consumption in a WNCS can be reduced and bandwidth can be more efficiently reused. To enable such scheduling, a model for the traffic generated by ETC is required. In \cite{kolarijani2016formal}, Kolarijani and Mazo propose a type of approximate power quotient system, to derive models that capture the timing behaviors of ETC systems applying the triggering mechanism from \cite{tabuada2007event}. They first partition the state space into finite cones. In each cone, they analyze the timing behavior by over-approximation methods (see e.g. \cite{cloosterman2010controller}, \cite{cloosterman2009stability}, \cite{donkers2011stability}, \cite{gielen2010polytopic}, \cite{hetel2006stabilization}, \cite{skaf2009analysis}, \cite{suh2008stability}), linear matrix inequality (LMI) methods, and reachability analysis (see e.g. \cite{chutinan1998computing} and \cite{chutinan2003computational}).

Similarly, in order to fully extract the potential gains from PETC with scheduling approaches, a model for the traffic generated by PETC is necessary. In this work, we present a construction of the timing models of the PETC implementations from \cite{heemels2013periodic}. First of all, we modify the PETC mechanism by giving an upper bound time such that if no event happens within that interval, the system will be forced to generate an event by the end of it. When constructing the models, the approach has two steps. We first divide the state space into a finite number of partitions. For a 2-dimensional system, the partition looks like a dartboard. Then we construct a set of LMIs to compute the output map. Transition relations among different regions are derived by computing the reachable state set starting from each region. Compared with the work from \cite{fiter2015robust}, we do not require that the perturbation should vanish as the state converges. Instead, we only assume the perturbation to be both $\mathcal{L}_2$ and $\mathcal{L}_{\infty}$.

This paper is organized as follows. Section \ref{section:notation} presents the necessary notation and definitions. The problem to be solved is defined in Section \ref{section:problemdefinition}. Section \ref{section:construction} shows all the details to construct a power quotient system to model the traffic of a centralized PETC implementation. A numerical example is shown in Section \ref{section:numericalexample}. Section \ref{section:conclusion} summarizes the contributions of this paper and discusses future work. To ease the readability, the proofs are collected in the Appendix.

\section{Notation and preliminaries}
\label{section:notation}

We denote the $n$-dimensional Euclidean space by $\mathbb{R}^n$, the positive real numbers by $\mathbb{R}^+$, by $\mathbb{R}^+_0=\mathbb{R}^+\cup\{0\}$. The natural numbers including zero is denoted by $\mathbb{N}$. When zero is not included, we denote the natural numbers as $\mathbb{N}^+$. $\mathbb{IN}^+$ is the set of all closed intervals $[a,b]$ such that $a,b\in\mathbb{N}^+$ and $a\leq b$. For any set $S$, $2^S$ denotes the set of all subsets of $S$, i.e. the power set of $S$. $\mathcal{M}_{m\times n}$ and $\mathcal{M}_n$ are the set of all $m\times n$ real valued matrices and the set of all $n\times n$ real-valued symmetric matrices respectively. A symmetric matrix $M\in \mathbb{R}^{n\times n}$ is said to be positive (negative) definite, denoted by $M\succ0$ ($M\prec0$), whenever $x^{\mathrm{T}}Mx>0$ ($x^{\mathrm{T}}Mx<0$) for all $x\neq 0$, $x\in\mathbb{R}^n$. $M\succeq 0$ ($M\preceq 0$) means $M$ is a positive (negative) semi-definite matrix. When $Q\subseteq Z\times Z$ is an equivalence relation on a set $Z$, $[z]$ denotes the equivalence class of $z\in Z$ and $Z/Q$ denotes the set of all equivalence classes. For a locally integrable signal $w$: $\mathbb{R}^{+}\rightarrow\mathbb{R}^n$, we denote by $\|w\|_{\mathcal{L}_{2}}=\sqrt{\int_0^{\infty}|w(t)|^2dt}$ its $\mathcal{L}_{2}$-norm, $\|w\|_{\mathcal{L}_{\infty}}=\sup_{t\geq 0} \|w(t)\|<\infty$ its $\mathcal{L}_{\infty}$-norm. Furthermore, we define the space of all locally integrable signals with a finite $\mathcal{L}_2$-norm as $\mathcal{L}_2$, the space of all signals with a finite $\mathcal{L}_{\infty}$-norm as $\mathcal{L}_{\infty}$.

Now we review some notions from the field of system theory.

\begin{definition}{(Metric)\cite{ewald2012combinatorial}}\label{definition:metric}
Consider a set $T$, $d:T\times T\rightarrow\mathbb{R}\cup\{+\infty\}$ is a metric (or a distance function) if the following three conditions are satisfied $\forall x,\,y,\,z\in T$:
\begin{itemize}
    \item $d(x,y)=d(y,x)$;
    \item $d(x,y)=0\leftrightarrow x=y$;
    \item $d(x,y)\leq d(x,z)+d(y,z)$.
\end{itemize}
The ordered pair $(T,d)$ is said to be a metric space.
\end{definition}

\begin{definition}{(Hausdorff distance)\cite{ewald2012combinatorial}}\label{definition:hausdorffdistance}
Assume $X$ and $Y$ are two non-empty subsets of a metric space $(T,d)$. The Hausdoorff distance $d_H(X,Y)$ is given by:
\begin{equation}\label{eq:hausdorffdistance}
\max\left\{\sup_{x\in X}\inf_{y\in Y}d(x,y),\sup_{y\in Y}\inf_{x\in X}d(x,y)\right\}.
\end{equation}
\end{definition}

\begin{definition}{(System)\cite{tabuada2009verification}}\label{difinition:system}
A system is a sextuple $(X,X_0,U,\longrightarrow,Y,H)$ consisting of:
\begin{itemize}
  \item a set of states $X$;
  \item a set of initial states $X_0\subseteq X$;
  \item a set of inputs $U$;
  \item a transition relation $\longrightarrow\subseteq X\times U\times X$;
  \item a set of outputs $Y$;
  \item an output map $H:X\rightarrow Y$.
\end{itemize}
\end{definition}
The term finite-state (infinite-state) system indicates $X$ is a finite (an infinite) set. For a system, if the cardinality of $U$ is smaller than or equal to one, then this system is said to be autonomous.

\begin{definition}{(Metric system)\cite{tabuada2009verification}}\label{definition:metricsystem}
A system $\mathcal{S}$ is said to be a metric system if the set of outputs $Y$ is equipped with a metric $d:Y\times Y\rightarrow\mathbb{R}_0^+$.
\end{definition}

\begin{definition}{(Approximate simulation relation)\cite{tabuada2009verification}}\label{definition:approximatesimulationrelation}
Consider two metric systems $\mathcal{S}_a$ and $\mathcal{S}_b$ with $Y_a=Y_b$, and let $\epsilon\in\mathbb{R}_0^+$. A relation $R\subseteq X_a\times X_b$ is an $\epsilon$-approximate simulation relation from $\mathcal{S}_a$ to $\mathcal{S}_b$ if the following three conditions are satisfied:
\begin{itemize}
    \item $\forall x_{a0}\in X_{a0}$, $\exists x_{b0}\in X_{b0}$ such that $(x_{a0},x_{b0})\in R$;
    \item $\forall (x_a,x_b)\in R$ we have $d\left(H_a(x_a),H_b(x_b)\right)\leq\epsilon$;
    \item $\forall (x_a,x_b)\in R$ such that $(x_a,u_a,x'_a)\in\underset{a}{\longrightarrow}$ in $\mathcal{S}_a$ implies $\exists(x_b,u_b,x'_b)\in\underset{b}{\longrightarrow}$ in $\mathcal{S}_b$ satisfying $(x'_a,x'_b)\in R$.
\end{itemize}
\end{definition}
We denote the existence of an $\epsilon$-approximate simulation relation from $\mathcal{S}_a$ to $\mathcal{S}_b$ by $\mathcal{S}_a\preceq^{\epsilon}_{\mathcal{S}}\mathcal{S}_b$, and say that $\mathcal{S}_b$ $\epsilon$-approximately simulates $\mathcal{S}_a$ or $\mathcal{S}_a$ is $\epsilon$-approximately simulated by $\mathcal{S}_b$. Whenever $\epsilon=0$, the inequality $d(H_a(x_a),H_b(x_b))\leq\epsilon$ implies $H_a(x_a)=H_b(x_b)$ and the resulting relation is called an (exact) simulation relation.

We introduce a notion of power quotient system and corresponding lemma for later analysis.

\begin{definition}{(Power quotient system)\cite{kolarijani2016formal}}\label{definition:powerquotientsystem}
Let $\mathcal{S}=(X,X_0,U,\longrightarrow,Y,H)$ be a system and $R$ be an equivalence relation on $X$. The power quotient of $\mathcal{S}$ by $R$, denoted by $\mathcal{S}_{/R}$, is the system $\left(X_{/R},X_{/R,0},U_{/R},\underset{/R}{\longrightarrow},Y_{/R},H_{/R}\right)$ consisting of:
\begin{itemize}
    \item $X_{/R}=X/R$;
    \item $X_{/R,0}=\left\{x_{/R}\in X_{/R}|x_{/R}\cap X_{0}\neq \emptyset\right\}$;
    \item $U_{/R}=U$;
    \item $\left(x_{/R},u,x'_{/R}\right)\in\underset{/R}{\longrightarrow}$ if $\exists(x,u,x')\in\longrightarrow$ in $\mathcal{S}$ with $x\in x_{/R}$ and $x'\in x'_{/R}$;
    \item $Y_{/R}\subset 2^{Y}$;
    \item $H_{/R}\left(x_{/R}\right)=\bigcup_{x\in x_{/R}}H(x)$.
\end{itemize}
\end{definition}

\begin{lemma}{[Lemma 1 in \cite{kolarijani2016formal}]}\label{lemma:approximatelysimulation}
Let $\mathcal{S}$ be a metric system, $R$ be an equivalence relation on $X$, and let the metric system $\mathcal{S}_{/R}$ be the power quotient system of $\mathcal{S}$ by $R$. For any
\begin{equation}
\epsilon\geq\max_{\begin{aligned}x&\in x_{/R}\\x_{/R}&\in X_{/R}\end{aligned}}\,d\left(H(x),H_{/R}\left(x_{/R}\right)\right),
\end{equation}
with $d$ the Hausdorff distance over the set $2^Y$, $\mathcal{S}_{/R}$ $\epsilon$-approximately simulates $\mathcal{S}$, i.e. $\mathcal{S}\preceq_{\mathcal{S}}^{\epsilon}\mathcal{S}_{/R}$.
\end{lemma}

The definition of Minkowski addition is introduced here for the computation of the reachable sets.

\begin{definition}{(Minkowski addition)}\label{definition:minkowskiaddition}
The Minkowski addition of two sets of vectors $\mathcal{A}$ and $\mathcal{B}$ in Euclidean space is formed by adding each vector in $\mathcal{A}$ to each vector in $\mathcal{B}$:
\begin{equation*}
\mathcal{A}\oplus\mathcal{B}=\{\textbf{a}+\textbf{b}|\textbf{a}\in \mathcal{A},\textbf{b}\in \mathcal{B}\},
\end{equation*}
where $\oplus$ denotes the Minkowski addition.
\end{definition}

\section{Problem definition}
\label{section:problemdefinition}

The centralized PETC presented in \cite{heemels2013periodic} is reviewed here. Consider a continuous linear time-invariant (LTI) plant of the form:
\begin{equation}\label{eq:system}
\left\{
\begin{aligned}
  \dot{\xi}_p(t)&=A_p\xi_p(t)+B_p\hat{v}(t)+Ew(t)\\
  y(t)&=C_p\xi_p(t),\\
\end{aligned}
\right.
\end{equation}
where $\xi_p(t)\in\mathbb{R}^{n_p}$ denotes the state vector of the plant, $y(t)\in\mathbb{R}^{n_y}$ denotes the plant output vector, $\hat{v}(t)\in\mathbb{R}^{n_v}$ denotes the input applied to the plant, $w(t)\in\mathbb{R}^{n_w}$ denotes the perturbation. The plant is controlled by a discrete-time controller, given by:
\begin{equation}\label{eq:controllaw}
\left\{\begin{aligned}
\xi_c(t_{k+1})&=A_c\xi_c(t_k)+B_c\hat{y}(t_k)\\
v(t_k)&=C_c\xi_c(t_k)+D_c\hat{y}(t_{k}),
\end{aligned}
\right.
\end{equation}
where $\xi_c(t_k)\in\mathbb{R}^{n_c}$ denotes the state vector of the controller, $v(t_k)\in\mathbb{R}^{n_v}$ denotes the controller output vector, and $\hat{y}(t_k)\in\mathbb{R}^{n_y}$ denotes the input applied to the controller. A periodic sampling sequence is given by:
\begin{equation}\label{eq:samplingsequence}
\mathcal{T}_s:=\{t_k|t_k:=kh,k\in\mathbb{N}\},
\end{equation}
where $h>0$ is the sampling interval. Define two vectors:
\begin{equation}\label{eq:output}
\begin{aligned}
u(t):&=\begin{bmatrix}
            y^{\mathrm{T}}(t) & v^{\mathrm{T}}(t) \\
          \end{bmatrix}^{\mathrm{T}}\in\mathbb{R}^{n_u}\\
\hat{u}(t_k):&=\begin{bmatrix}
            \hat{y}^{\mathrm{T}}(t_k) & \hat{v}^{\mathrm{T}}(t_k)  \\
          \end{bmatrix}^{\mathrm{T}}\in\mathbb{R}^{n_u},
\end{aligned}
\end{equation}
with $n_u:=n_y+n_v$. $u(t)$ is the output of the implementation, $\hat{u}(t)$ is the input of the implementation. A zero-order hold mechanism is applied between samplings to the input. At each sampling time $t_k$, the input applied to the implementation $\hat{u}(t_k)$ is updated $\forall t_k\in\mathcal{T}_s$:
\begin{equation}\label{eq:sampleandupdate}
\hat{u}(t_k)=\left\{\begin{aligned}
&u(t_k),&\,&\text{if }\|u(t_k)-\hat{u}(t_k)\|>\sigma\|u(t_k)\|\\
&\hat{u}(t_{k-1}),&\,&\text{if }\|u(t_k)-\hat{u}(t_k)\|\leq\sigma\|u(t_k)\|,
\end{aligned}\right.
\end{equation}
where $\sigma>0$ is a given constant. Reformulating the event condition as a quadratic form, the event sequence can be defined by:
\begin{equation}\label{eq:eventsequence}
\mathcal{T}_e:=\left\{t_b|b\in\mathbb{N},t_b\in\mathcal{T}_s,\xi^{\mathrm{T}}(t_b)Q\xi(t_b)>0\right\}.
\end{equation}
where $\xi(t):=\begin{bmatrix}
\xi_p^{\mathrm{T}}(t) & \xi_c^{\mathrm{T}}(t) &
\hat{y}^{\mathrm{T}}(t) & \hat{v}^{\mathrm{T}}(t) \\
\end{bmatrix}^{\mathrm{T}}\in\mathbb{R}^{n_{\xi}}$, with $n_{\xi}:=n_p+n_c+n_y+n_v$. And:
\begin{equation*}
\begin{aligned}
Q=\begin{bmatrix}
Q_1 & Q_2 \\
Q_2^{\mathrm{T}} & Q_4 \\
\end{bmatrix}
\end{aligned}
\end{equation*}
in which:
\begin{equation*}\left\{
\begin{aligned}
Q_1&=\begin{bmatrix}
  (1-\sigma)C_p^{\mathrm{T}}C_p & \mathbf{0} \\
  \mathbf{0} & (1-\sigma)C_c^{\mathrm{T}}C_c \\
\end{bmatrix}\\
Q_2&=\begin{bmatrix}
       -C_p^{\mathrm{T}} & \mathbf{0} \\
       (1-\sigma)C_c^{\mathrm{T}}D_c & -C_c^{\mathrm{T}} \\
     \end{bmatrix}\\
Q_4&=\begin{bmatrix}
       I+(1-\sigma)D_c^{\mathrm{T}}D_c & -D_c^{\mathrm{T}} \\
       -D_c & I \\
     \end{bmatrix}
\end{aligned}\right.
\end{equation*}
$\mathbf{0}$ is a zero matrix with proper dimension, $I$ is an identity matrix with appropriate dimension.

It is obvious that $\mathcal{T}_e\subseteq\mathcal{T}_s$. According to Theorem V.2 in \cite{heemels2013periodic}, if the hypothesis therein are satisfied, then the system (\ref{eq:system}-\ref{eq:eventsequence}):
\begin{enumerate}
  \item is globally exponential stable (GES), i.e. $\exists c>0$ and $\rho>0$ s.t. $\forall\xi(0)\in\mathbb{R}^{n_{\xi}}$ with $w=0$, $\|\xi(t)\|\leq ce^{-\rho t}\|\xi(0)\|$ for all $t\in\mathbb{R}^+$.
  \item has $\mathcal{L}_2$-gain from $w$ to $z$ smaller than or equal to $\gamma$, i.e. $\exists\sigma:\mathbb{R}^{n_{\xi}}\rightarrow\mathbb{R}^+$ s.t. $\forall w\in\mathcal{L}_2$, $\xi(0)\in\mathbb{R}^{n_{\xi}}$, the corresponding solution to the system and $z(t):=g(\xi(t),w(t))$ satisfies $\|z\|_{\mathcal{L}_2}\leq\sigma(\xi(0))+\gamma\|w\|_{\mathcal{L}_2}$.
\end{enumerate}

To model the timing behaviour of a PETC system, we aim at constructing a power quotient system for this implementation.

\begin{remark}\label{remark:noupperbound}
Because of the uncertainty brought by the perturbation, it may happen that the perturbation compensates the effect of sampling, helping the state of the implementation to converge. Therefore the event condition in (\ref{eq:eventsequence}) may not be satisfied along the timeline. As a result, there may not be an upper bound for the event intervals. However an upper bound is necessary for constructing a useful power quotient system.
\end{remark}

\begin{remark}\label{remark:observability}
To apply scheduling approaches, an online scheduler is required. The model we are going to construct is non-deterministic, meaning that after an event the system may end up in several possible regions, but those regions are defined in terms of $\xi_p$, which means that from a measurement is not always clear in which region the system is. That means this online scheduler cannot figure out where the system is from simple output measurements. Therefore, the online scheduler should be able to access in which region the system is.
\end{remark}

\begin{assumption}\label{assumption:observability}
The current state region at each event-triggered time $t_b$ can be obtained in real time.
\end{assumption}

Because of the observation in Remark \ref{remark:noupperbound}, we use instead the following event condition:
\begin{equation}\label{eq:newcondition}
\begin{aligned}
t_{b+1}=&\inf\left\{t_{k}\left|t_k\in\mathcal{T}_s,t_k>t_b, \right.\right.\\ &\left.\xi^{\mathrm{T}}(t_k)Q\xi(t_k)>0\bigvee t_{k}\geq t_{b}+\bar\tau_{\mathcal{R}(\xi(t_b))}\right\},
\end{aligned}
\end{equation}
where $\mathcal{R}(\xi(t_b))$ is the state region on state-space $\mathbb{R}^{n_{\xi}}$ at last sampling time $t_b$, $\bar\tau_{\mathcal{R}(\xi(t_b))}$ is a regional maximum allowable event interval (MAEI), which is dependent on $\mathcal{R}(\xi(t_b))$. According to Assumption \ref{assumption:observability}, $\mathcal{R}(\xi(t_b))$ is obtainable. If this value is not possible to be accessed by the triggering mechanisms, one can always employ a global upper bound $\bar\tau:\geq\bar\tau_{\mathcal{R}(\xi(t_b))}$. We will discuss the computation of $\bar\tau_{\mathcal{R}(t_b)}$ in later sections. Note that, if the PETC implementation employing (\ref{eq:eventsequence}) can guarantee some pre-designed stability and performance, then the PETC implementation employing (\ref{eq:newcondition}) can guarantee the same stability and performance.

Consider a period:
\begin{equation}\label{eq:tauxkx}
\tau(x):=t_{b+1}-t_b=k_xh.
\end{equation}
By definition $\hat{u}(t)$ is constant $\forall t\in[t_b,t_{b+1}[$ and dependent on $\xi_p(t_b)$ and $\xi_c(t_b)$. The input $\hat{u}(t)$ can be expressed as:
\begin{equation*}
\hat{u}(t)=C_Ex,\,C_E:=
\begin{bmatrix}
  C_p & \mathbf{0} \\
  D_cC_p & C_c \\
\end{bmatrix},
\end{equation*}
where
\begin{equation*}
x:=\begin{bmatrix}
\xi_p^{\mathrm{T}}(t_b) & \xi_c^{\mathrm{T}}(t_b) \\
\end{bmatrix}^{\mathrm{T}}.
\end{equation*}
Let $\xi_x(k):=\begin{bmatrix}
\xi_p^{\mathrm{T}}(t_b+kh) & \xi_c^{\mathrm{T}}(t_b+kh) \\
\end{bmatrix}^{\mathrm{T}}$ be the state evolution with initial state $x=\begin{bmatrix}
\xi_p^{\mathrm{T}}(t_b) & \xi_c^{\mathrm{T}}(t_b) \\
\end{bmatrix}^{\mathrm{T}}$, and $k\in\mathbb{N}$. Now $\xi_x(k)$ can be computed as:
\begin{equation}\label{eq:relativestate}
\xi_x(k)=M(k)x+\Theta(k),
\end{equation}
where
\begin{equation*}
M(k):=\begin{bmatrix}
        M_1(k) \\
        M_2(k) \\
      \end{bmatrix},\,
\Theta(k):=\begin{bmatrix}
        \Theta_1(k) \\
        \mathbf{0} \\
      \end{bmatrix},
\end{equation*}
\begin{equation*}
\left\{\begin{aligned}
M_1(k):=&\begin{bmatrix}
        I & \mathbf{0} \\
      \end{bmatrix}+\int_0^{kh}e^{A_ps}ds\left(A_p\begin{bmatrix}
        I & \mathbf{0} \\
      \end{bmatrix}\right.\\
      &\left.+B_p\begin{bmatrix}
        D_cC_p & C_c \\
      \end{bmatrix}\right),\\
M_2(k):=&A_c^k\begin{bmatrix}
        \mathbf{0} & I \\
      \end{bmatrix}+\sum_{i=0}^{k-1}A_c^{k-1-i}B_c\begin{bmatrix}
        C_p & \mathbf{0} \\
      \end{bmatrix},\\
\Theta_1(k):=&\int_{0}^{kh}e^{A_p(kh-s)}Ew(s)ds.
\end{aligned}\right.
\end{equation*}
Thus from the event condition in (\ref{eq:newcondition}), $k_x$ in (\ref{eq:tauxkx}) can be computed by:
\begin{equation}\label{eq:xeventsequence}
\begin{aligned}
k_x=\min\left\{\underline{k}_x,\overline{k}_x\right\},
\end{aligned}
\end{equation}
where $\overline{k}_x:=\frac{\bar\tau_{\mathcal{R}(x)}}{h}$ and
\begin{equation}\label{eq:underlinekx}
\begin{aligned}
\underline{k}_x:=&\inf\left\{k\in\mathbb{N}^+\right|\\
&\left.\left.\begin{bmatrix}
  M(k)x+\Theta(k) \\
    C_Ex \\
\end{bmatrix}^{\mathrm{T}}Q
\begin{bmatrix}
  M(k)x+\Theta(k) \\
  C_Ex \\
\end{bmatrix}>0\right\}\right\}.
\end{aligned}
\end{equation}

Now we present the main problem to be solved in this paper. Consider the system:
\begin{equation}\label{eq:Asystem}
\mathcal{S}=(X,X_0,U,\longrightarrow,Y,H),
\end{equation}
where
\begin{itemize}
  \item $X=\mathbb{R}^{n_{x}}$, $n_{x}=n_{p}+n_{c}$;
  \item $X_0\subseteq \mathbb{R}^{n_{x}}$;
  \item $U=\emptyset$;
  \item $\longrightarrow\subseteq X\times U\times X$ such that $\forall x,x'\in X:(x,x')\in\longrightarrow$ iff $\xi_x(H(x))=x'$;
  \item $Y\subset\mathbb{N}^+$;
  \item $H:\mathbb{R}^{n_{x}}\rightarrow \mathbb{N}^+$ where $H(x)=k_x$.
\end{itemize}
$\mathcal{S}$ is an infinite-state system. The output set $Y$ of system $\mathcal{S}$ contains all the possible amount of sampling steps $\frac{t_{b+1}-t_b}{h}\in\mathbb{N}$, $b\in\mathbb{N}$ that the system (\ref{eq:system}-\ref{eq:sampleandupdate}), and (\ref{eq:newcondition}) may exhibit. Once the sampling time $h$ is chosen, the event interval then can be computed by $k_xh$.

\begin{problem}\label{problem:eventsequence}
Construct a finite abstraction of system $\mathcal{S}$ capturing enough information for scheduling.
\end{problem}

Inspired by \cite{kolarijani2016formal}, we solve this problem by constructing a power quotient systems $\mathcal{S}_{/P}$ based on an adequately designed equivalence relation $P$ defined over the state set $X$ of $\mathcal{S}$. The constructed systems $\mathcal{S}_{/P}$ are semantically equivalent to timed automata, which can be used for automatic scheduler design \cite{kolarijani2015symbolic}.

In particular, the system $\mathcal{S}_{/P}$ to be constructed is as follows:
\begin{equation}\label{eq:Psystem}
\mathcal{S}_{/P}=\left(X_{/P},X_{/P,0},U_{/P},\underset{/P}{\longrightarrow},Y_{/P},H_{/P}\right),
\end{equation}
\begin{itemize}
    \item $X_{/P}=\mathbb{R}^{n_x}_{/P}:=\{\mathcal{R}_1,\cdots,\mathcal{R}_q\}$;
    \item $X_{/P,0}=\mathbb{R}^{n_{x}}_{/P}$;
    \item $\left(x_{/P},x'_{/P}\right)\in\underset{/P}{\longrightarrow}$ if $\exists x\in x_{/P}$, $\exists x'\in x'_{/P}$ such that $\xi_x(H(x))=x'$;
    \item $Y_{/P}\subset 2^{Y}\subset\mathbb{IN}^+$;
    \item $H_{/P}\left(x_{/P}\right)=\left[\min_{x\in x_{/P}} H(x),\max_{x\in x_{/P}} H(x)\right]:=\left[\underline{k}_{x_{/P}},\bar{k}_{x_{/P}}\right]$.
\end{itemize}
$\mathcal{S}_{/P}$ is a finite state system.

Compared with the power quotient system constructed in \cite{kolarijani2016formal}, a main difference is that since we focus on PETC, there is no timing uncertainty.

\section{Construction of the quotient system}
\label{section:construction}

\subsection{State set}

From the results in \cite{fiter2012state}, we remark the following fact:

\begin{remark}\label{remark:weq0}
When $w=0$, excluding the origin, all the states that lie on a line going through the origin have an identical triggering behavior.
\end{remark}

We also call the following assumption:

\begin{assumption}\label{assumption:w}
The perturbation $w$ satisfies $w\in\mathcal{L}_2$ and $w\in\mathcal{L}_{\infty}$. Besides, assume an upper bound $\mathcal{W}>0$ for $\|w\|_{\mathcal{L}_{\infty}}$, i.e. $\|w\|_{\mathcal{L}_{\infty}}\leq \mathcal{W}$, is known.
\end{assumption}

Base on Remark \ref{remark:weq0} and Assumption \ref{assumption:w}, we propose state-space partition as follows:
\begin{equation}\label{eq:stateset}
\begin{aligned}
\mathcal{R}_{s_1,s_2}=&\left\{x\in\mathbb{R}^{n_x}\left|\bigwedge_{i=1}^{n_x-1} x^{\mathrm{T}}\Xi_{s_1,(i,i+1)}x\geq 0\right.\right.\\
&\left.\bigwedge W_{s_2-1}\leq |x|< W_{s_2}\right\},\\
\end{aligned}
\end{equation}
where $s_1\in\{1,\cdots,q_{1}\}$, $s_2\in\{1,\cdots,q_{2}\}$, $q_1,\,q_2\in\mathbb{N}$ are pre-designed scalars. $\Xi_{s_1,(i,j)}$ is a constructed matrix; $\{W_{i}|i\in\{0,\cdots,q_2\}\}$ is a sequence of scalars. Note that $W_0=0$, $W_{q_2}=+\infty$, and the rest $W_{s_2}$ are bounded and somewhere in between $0$ and $+\infty$. It is obvious that $\bigcup_{s_1\in\{1,\cdots,q_1\},s_2\in\{1,\cdots,q_2\}}\mathcal{R}_{s_1,s_2}=\mathbb{R}^{n_x}$.

This state-space partition combines partitioning the state-space into finite polyhedral cones (named as \emph{isotropic covering} \cite{fiter2012state}) and finite homocentric spheres. From (\ref{eq:stateset}), we can see that, the \emph{isotropic covering} describes the relation between entries of the state vector, while the \emph{transverse isotropic covering} is used to capture the relation between the norm of the state vector and the $\mathcal{L}_{\infty}$ norm of the perturbations, which will be shown later in Theorem \ref{theorem:down}. If $w=0$, the homocentric spheres can be omitted.

Details on the \emph{isotropic covering} can be found in the Appendix. Figure \ref{fig:stateset} shows a 2-dimensional example.

\begin {figure}[!t]
\centering
\includegraphics[width=\linewidth]{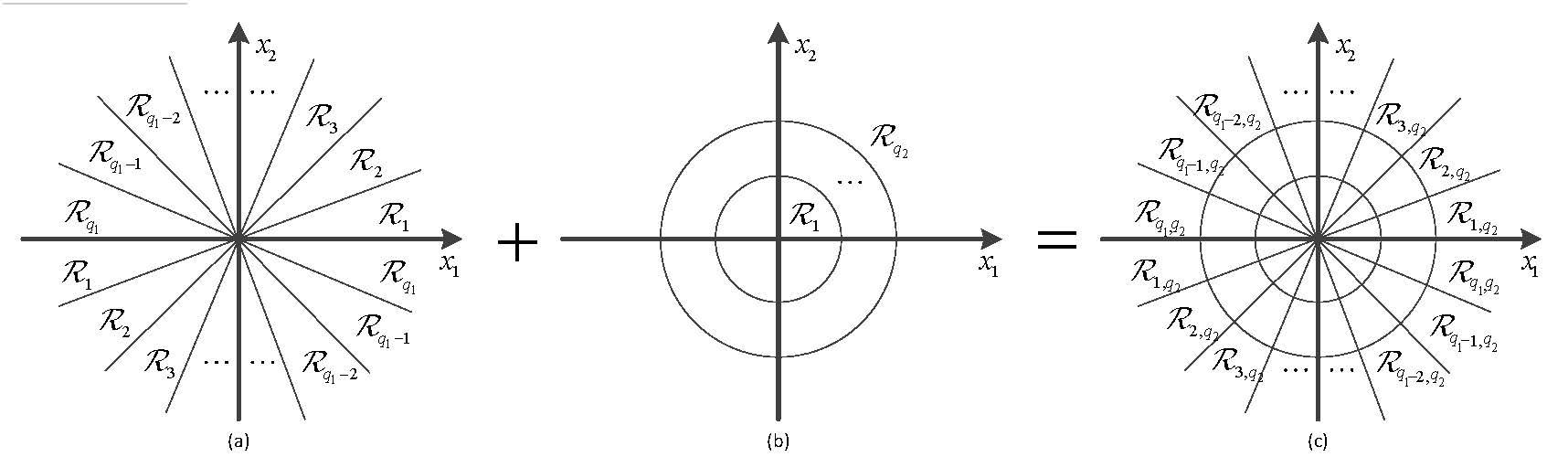}
\caption{An example of the state space partition, into (a) finite polyhedral cones, (b) finite homocentric spheres, and (c) finite state-space partition.}
\label{fig:stateset}
\end {figure}

\subsection{Output map}

We first free the system dynamics from the uncertainty brought by the perturbation.

\begin{lemma}\label{lemma:phiapproximate}
Consider the system (\ref{eq:system}-\ref{eq:sampleandupdate}) and (\ref{eq:newcondition}), and that Assumptions \ref{assumption:w} hold. If there exist a scalar $\mu\geq 0$ and a symmetric matrix $\Psi$ such that $(Q_1+\Psi)_1\preceq\mu I$, then $\underline{k}_x$ generated by (\ref{eq:underlinekx}) is lower bounded by:
\begin{equation}\label{eq:phiapproximate}
k_x':=\inf\{k\in\mathbb{N}^{+}|\Phi(k)\succ0\},
\end{equation}
where
\begin{equation*}
\begin{aligned}
&Q_1+\Psi=\begin{bmatrix}
        (Q_1+\Psi)_1 & (Q_1+\Psi)_2 \\
        (Q_1+\Psi)_3 & (Q_1+\Psi)_4 \\
\end{bmatrix}\\
&(Q_1+\Psi)_1\in\mathbb{R}^{n_p\times n_p},
\end{aligned}
\end{equation*}
\begin{equation}\label{eq:Ephi}
\Phi(k):=\begin{bmatrix}
            \Phi_{1}(k) & \Phi_2(k) & \mathbf{0} \\
            \Phi_2^{\mathrm{T}}(k) & -\Psi & \mathbf{0} \\
            \mathbf{0} & \mathbf{0} &  \Phi_{3}(k)\\
          \end{bmatrix},
\end{equation}
\begin{equation}\label{eq:Phi123}
\left\{
\begin{aligned}
\Phi_{1}(k)=&M^{\mathrm{T}}(k)Q_1M(k)+M^{\mathrm{T}}(k)Q_2C_E\\
&+C_E^{\mathrm{T}}Q_3M(k)+ C_E^{\mathrm{T}}Q_4C_E\\
\Phi_{2}(k)=&M^{\mathrm{T}}(k)Q_1+C_E^{\mathrm{T}}Q_3\\
\Phi_{3}(k)=&kh\mu\lambda_{\max}\left(E^{\mathrm{T}}E\right)d_{A_p}(k),
\end{aligned}
\right.
\end{equation}
and
\begin{equation*}
d_{A_p}(k)=\left\{
\begin{aligned}
&\frac{e^{k\lambda_{\max}\left(A_p+A_p^{\mathrm{T}}\right)}-1} {\lambda_{\max}\left(A_p+A_p^{\mathrm{T}}\right)},&\text{if }\lambda_{\max}\left(A_p+A_p^{\mathrm{T}}\right)\neq 0,\\
&kh,&\text{if }\lambda_{\max}\left(A_p+A_p^{\mathrm{T}}\right)=0.
\end{aligned}\right.
\end{equation*}
\end{lemma}

Next we construct LMIs that bridge Lemma \ref{lemma:phiapproximate} and the state-space partition.

\begin{theorem}{(Regional lower bound)}\label{theorem:down}
Consider a scalar $\underline{k}_{s_1,s_2}\in\mathbb{N}$ and regions with $s_2>1$. If all the hypothesis in Lemma \ref{lemma:phiapproximate} hold and there exist scalars $\underline{\varepsilon}_{k,(s_1,s_2),(i,i+1)}\geq 0$ where $i\in\{1,\cdots,n_x-1\}$ such that for all $k\in\{0,\cdots,\underline{k}_{s_1,s_2}\}$ the following LMIs hold:
\begin{equation}\label{eq:mainLMI}
\begin{aligned}
\begin{bmatrix}
 H & \Phi_2(k) \\
 \Phi_2^{\mathrm{T}}(k) & -\Psi \\
\end{bmatrix}\preceq 0,\\
\end{aligned}
\end{equation}
where
\begin{equation*}
\begin{aligned}
H=&\Phi_1(k)+\Phi_3(k)\mathcal{W}^2W_{s_2-1}^{-2}I\\
&+\sum_{i\in\{1,\cdots,n_x-1\}}\underline\varepsilon_{k,(s_1,s_2),(i,i+1)}\Xi_{s_1,(i,i+1)},
\end{aligned}
\end{equation*}
with $\Phi_1(k)$, $\Phi_2(k)$, and $\Phi_3(k)$ defined in (\ref{eq:Phi123}), and $\Psi$ from Lemma \ref{lemma:phiapproximate}, then the inter event times (\ref{eq:newcondition}) for system (\ref{eq:system}-\ref{eq:sampleandupdate}) are regionally bounded from below by $(\underline{k}_{s_1,s_2}+1)h$. For the regions with $s_2=1$, the regional lower bound is $h$.
\end{theorem}

\begin{remark}
In Theorem \ref{theorem:down}, we discuss the situations when $s_2>1$ and $s_2=1$, since for all regions with $s_2>1$, it holds that $W_{s_2-1}\neq 0$; while for all regions with $s_2=1$, $W_{s_2-1}=0$ holds. When $W_{s_2-1}\neq 0$, one can easily validate the feasibility of the LMI (\ref{eq:mainLMI}); while when $W_{s_2-1}=0$, $H$ will be diagonal infinity, making the LMI (\ref{eq:mainLMI}) infeasible when $k>0$. However, according to the property of PETC, i.e. $t_{b+1}\in\mathcal{T}_s$ and $t_{b+1}>t_{b}$, the regional lower bound exists and is equal to $h$.
\end{remark}

Following similar ideas of Theorem \ref{theorem:down}, we present next lower and upper bounds starting from each state partition when $w=0$.  Consider the following event condition:
\begin{equation}\label{eq:nonxeventsequence}
\begin{aligned}
&k_{x}=\inf\left\{k\in\mathbb{N}^+\left|\begin{bmatrix}
                                     M(k)x \\
                                     C_Ex \\
                                   \end{bmatrix}^{\mathrm{T}}Q
                                   \begin{bmatrix}
                                     M(k)x \\
                                     C_Ex \\
                                   \end{bmatrix}>0\right.\right\}.
\end{aligned}
\end{equation}

\begin{remark}\label{remark:upperbound}
Since (\ref{eq:nonxeventsequence}) does not consider perturbations, when computing the lower and upper bound for each region, according to Remark \ref{remark:weq0}, only applying \emph{isotropic covering} is enough.
\end{remark}

We define $\mathcal{R}_{s_1,\bullet}$ to represent $\mathcal{R}_{s_1,s_2}$, $\forall s_2\in\{1,\cdots,q_2\}$.

\begin{corollary}
{(Regional lower bound when $w=0$)}\label{corollary:downw0}
Consider a scalar $\underline{k}_{s_1,\bullet}\in\mathbb{N}$. If there exist scalars $\underline{\varepsilon}_{k,s_1,(i,i+1)}\geq 0$ where $i\in\{1,\cdots,n_x-1\}$ such that for all $k\in\{0,\cdots,\underline{k}_{s_1,\bullet}\}$ the following LMIs hold:
\begin{equation}\label{eq:corollaryLMI}
\begin{aligned}
\Phi_1(k)+\sum_{i\in\{1,\cdots,n_x-1\}}\underline\varepsilon_{k,s_1,(i,i+1)}\Xi_{s_1,(i,i+1)}\preceq 0,
\end{aligned}
\end{equation}
with $\Phi_1(k)$ defined in (\ref{eq:Phi123}), then the inter event time (\ref{eq:eventsequence}) of the system (\ref{eq:system}-\ref{eq:sampleandupdate}) with $w=0$ are regionally bounded from below by $(\underline{k}_{s_1,\bullet}+1)h$.
\end{corollary}

\begin{corollary}{(Regional upper bound when $w=0$)}\label{theorem:up}
Let $\bar{l}\in\mathbb{N}$ be a large enough scalar. Consider a scalar $\bar{k}_{s_1,\bullet}\in\left\{\underline{k}_{s_1,\bullet},\cdots,\bar{l}\right\}$. If there exist scalars $\bar{\varepsilon}_{k,s_1,(i,i+1)}\geq 0$ where $i\in\{1,\cdots,n_x-1\}$ such that for all $k\in\left\{\bar{k}_{s_1,\bullet},\cdots,\bar{l}\right\}$ the following LMIs hold:
\begin{equation}\label{eq:upperLMI}
\begin{aligned}
\Phi_1(k)-\sum_{i\in\{1,\cdots,n_x-1\}}\bar\varepsilon_{k,s_1,(i,i+1)}\Xi_{s_1,(i,i+1)}\succ 0,
\end{aligned}
\end{equation}
with $\Phi_1(k)$ defined in (\ref{eq:Phi123}), then the inter event time (\ref{eq:eventsequence}) of the system (\ref{eq:system}-\ref{eq:sampleandupdate}) with $w=0$ are regionally bounded from above by $\bar{k}_{s_1,\bullet}h$.
\end{corollary}

\begin{remark}
For the choice of $\bar{l}$, we follow Remark 2 in \cite{kolarijani2016formal}, and apply a line search approach: increasing $\bar{l}$ until $\Phi_1(\bar{l})\succ 0$. This results in $\bar{l}$ being a global upper bound for the inter event time (\ref{eq:eventsequence}) of the system (\ref{eq:system}-\ref{eq:sampleandupdate}) with $w=0$.
\end{remark}

It is obvious that $\bar{l}\geq\bar{k}_{s_1,\bullet}>\underline{k}_{s_1,\bullet}\geq\underline{k}_{s_1,s_2}$, $\forall s_2$. We can now set the regional MAEI $\bar\tau_{\mathcal{R}(\xi(t_b))}$ in (\ref{eq:newcondition}) as: $\bar\tau_{\mathcal{R}(\xi(t_b))}:=\bar{k}_{s_1,\bullet}h$, $\forall x\in\mathcal{R}_{s_1,\bullet}$.

\subsection{Transition relation}

In this subsection, we discuss the construction of the transition relation and the reachable state set. Denote the initial state set as $X_{0,(s_1,s_2)}$, after $k$-th samplings without an update, the reachable state set is denoted as $X_{k,(s_1,s_2)}$. According to (\ref{eq:relativestate}), a relation can be obtained as:
\begin{equation}\label{eq:relativestateset}
X_{k,(s_1,s_2)}=M(k)X_{0,(s_1,s_2)}+\Theta(k).
\end{equation}
It is obvious that, $X_{k,(s_1,s_2)}$ cannot be computed directly, because the perturbation is uncertain and the state region may not be convex. Therefore, we aim to find sets $\hat{X}_{k,(s_1,s_2)}$ such that:
\begin{equation*}
X_{k,(s_1,s_2)}\subseteq\hat{X}_{k,(s_1,s_2)}.
\end{equation*}
To compute $\hat{X}_{k,(s_1,s_2)}$, we take the following steps:

\subsubsection{Partition the dynamics}

According to (\ref{eq:relativestateset}), $\hat{X}_{k,(s_1,s_2)}$ can be computed by:
\begin{equation*}
\hat{X}_{k,(s_1,s_2)}=\hat{X}^1_{k,(s_1,s_2)}\oplus\hat{X}^2_{k,(s_1,s_2)},
\end{equation*}
where $\oplus$ is the Minkowski addition, $\hat{X}^1_{k,(s_1,s_2)}$ and $\hat{X}^2_{k,(s_1,s_2)}$ are sets to be computed.

\subsubsection{Compute $\hat{X}^1_{k,(s_1,s_2)}$}

One can compute $\hat{X}^1_{k,(s_1,s_2)}$ by:
\begin{equation*}
\begin{aligned}
\hat{X}^1_{k,(s_1,s_2)}=M(k)\hat{X}_{0,(s_1,s_2)},
\end{aligned}
\end{equation*}
where $\hat{X}_{0,(s_1,s_2)}$ is a polytope that over approximates $X_{0,(s_1,s_2)}$, i.e. $X_{0,(s_1,s_2)}\subseteq\hat{X}_{0,(s_1,s_2)}$. $\hat{X}_{0,(s_1,s_2)}$ can be computed as in the optimization problem (1) in \cite{chutinan2003computational}.

\subsubsection{Compute $\hat{X}^2_{k,(s_1,s_2)}$}

For the computation of $\hat{X}^2_{k,(s_1,s_2)}$, it follows that:
\begin{equation*}
\hat{X}^2_{k,(s_1,s_2)}=\{x\in\mathbb{R}^{n_x}||x|\leq|\Theta(k)|\},
\end{equation*}
where
\begin{equation*}
\begin{aligned}
|\Theta(k)|&=\left|\int_{0}^{kh}e^{A_p(kh-s)}Ew(s)ds\right|\\
&\leq\int_{0}^{kh}\left|e^{A_p(kh-s)}Ew(s)\right|ds\\
&\leq\int_0^{kh}\left|e^{A_p(kh-s)}\right|ds|E|\|w\|_{\mathcal{L}_\infty}\\
&\leq\int_0^{kh}e^{\lambda_{\max}\left(\frac{A_p^{\mathrm{T}}+A_p}{2}\right)(kh-s)}ds|E|\mathcal{W}.
\end{aligned}
\end{equation*}
In which the last inequation holds according to (2.2) in \cite{van1977sensitivity}.

Thus the reachable set $X_{\left\{\underline{k}_{s_1,s_2},\overline{k}_{s_1,\bullet}\right\},(s_1,s_2)}$ of the system (\ref{eq:system}-\ref{eq:sampleandupdate}), and (\ref{eq:newcondition}) starting from $X_{0,(s_1,s_2)}$ can be computed by:
\begin{equation*}
\begin{aligned}
X_{\left\{\underline{k}_{s_1,s_2},\overline{k}_{s_1,\bullet}\right\},(s_1,s_2)}&\subseteq \hat{X}_{\left\{\underline{k}_{s_1,s_2},\overline{k}_{s_1,\bullet}\right\},(s_1,s_2)}\\
&=\bigcup_{k\in\left\{\underline{k}_{s_1,s_2},\cdots,\overline{k}_{s_1,\bullet}\right\}}\hat{X}_{k,(s_1,s_2)}.
\end{aligned}
\end{equation*}

To compute the transitions in $\mathcal{S}_{/P}$, one can check the intersection between the over approximation of reachable state set and all the state regions $\mathcal{R}_{s_1',s_2'}$, $\forall s_1'\in\{1,\cdots,q_1\},s_2'\in\{1,\cdots,q_2\}$. More specifically, one can check if the following feasibility problem for each state region holds:
\begin{equation*}
\begin{aligned}
\mathcal{R}_{s_1',s_2'}\cap\hat{X}_{\left\{\underline{k}_{s_1,s_2},\overline{k}_{s_1,\bullet}\right\},(s_1,s_2)}\neq\emptyset,\\
\end{aligned}
\end{equation*}
in which case
\begin{equation*}
\left(\mathcal{R}_{s_1,s_2},\mathcal{R}_{s_1',s_2'}\right)\in\underset{/P}{\longrightarrow}.
\end{equation*}

\subsection{Main result}

Now we summarize the main result of the paper in the following theorem.
\begin{theorem}\label{theorem:mainresult}
The metric system $\mathcal{S}_{/P}=\left(X_{/P},X_{/P,0},U_{/P}, \underset{/P}{\longrightarrow},Y_{/P},H_{/P}\right)$, $\epsilon$-approximately simulates $\mathcal{S}$, where $\epsilon=\max d_{H}\left(y,y'\right)$, $y=H(x)\in Y$, $y'=H_{/P}\left(x'\right)\in Y_{/P}$, $\forall \left(x,x'\right)\in P$, and $d_H(\cdot,\cdot)$ is the Hausdorff distance.
\end{theorem}

\section{Numerical example}
\label{section:numericalexample}

\begin {figure}[!t]
\centering
\includegraphics[width=\linewidth]{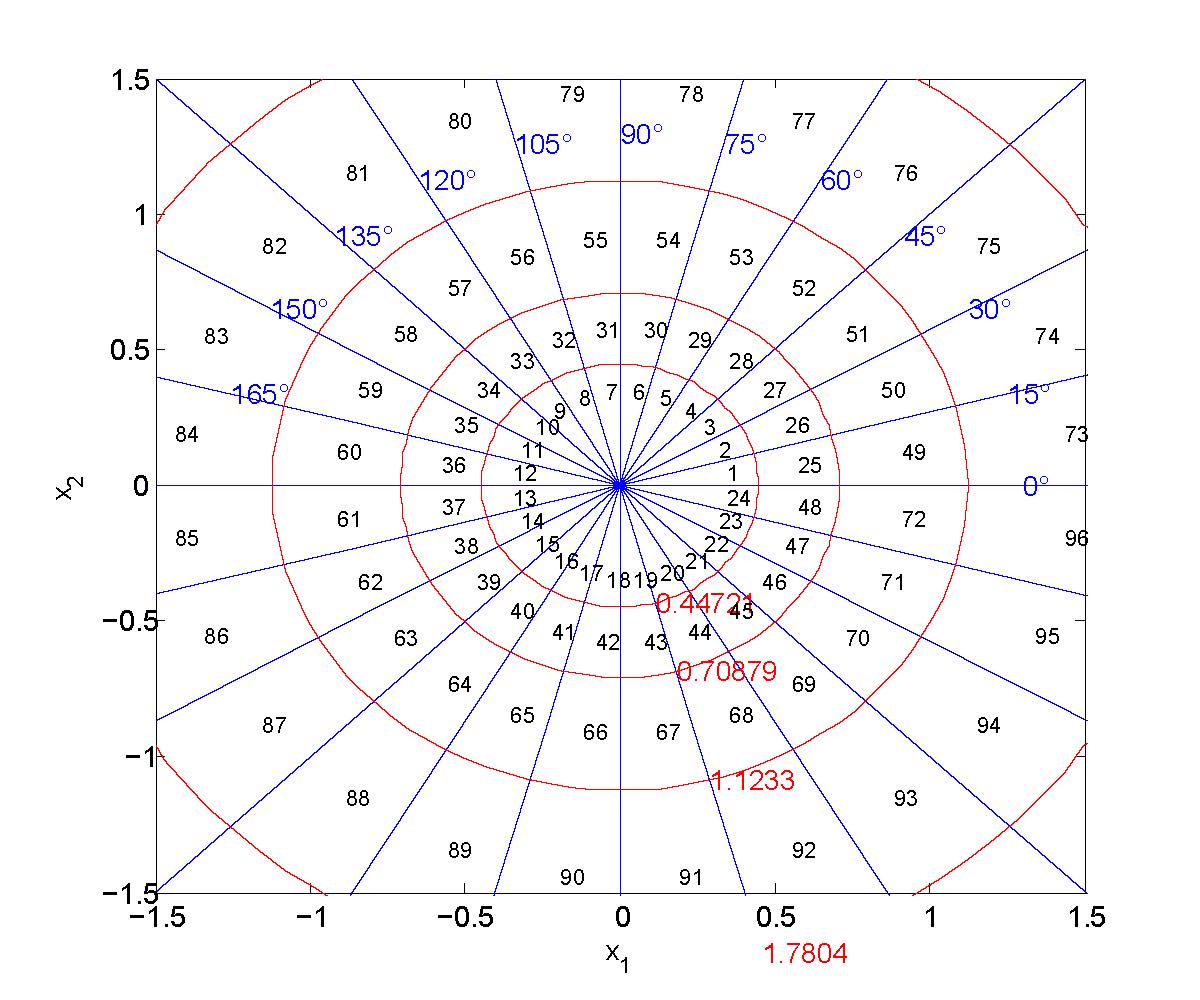}
\caption{State-space partition and the labeling of each region.}
\label{fig:gridpartition}
\end {figure}

In this section, we consider a system employed in \cite{heemels2013periodic} and \cite{tabuada2007event}. The plant is given by:
\begin{equation*}
\dot{\xi}(t)=\begin{bmatrix}
               0 & 1 \\
               -2 & 3 \\
             \end{bmatrix}\xi(t)+
             \begin{bmatrix}
               0 \\
               1 \\
             \end{bmatrix}v(t)+
             \begin{bmatrix}
               1 \\
               0 \\
             \end{bmatrix}w(t),
\end{equation*}
and the controller is given by:
\begin{equation*}
K=\begin{bmatrix}
    1 & -4 \\
  \end{bmatrix}.
\end{equation*}
This plant is chosen since it is easy to show the feasibility of the presented theory in 2-dimensional plots. The state-space partition is presented in Figure \ref{fig:gridpartition}.

\begin {figure}[!t]
\centering
\includegraphics[width=\linewidth]{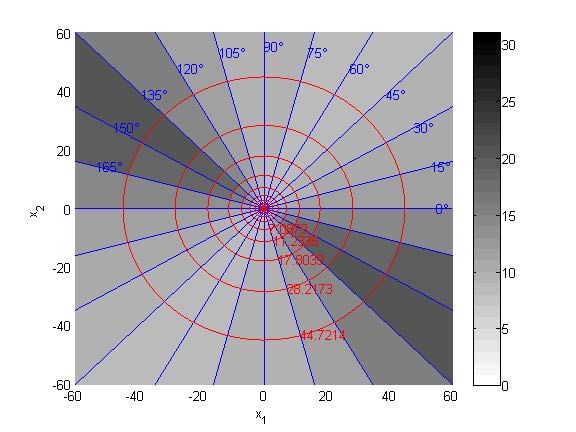}
\caption{Computed result of the regional lower bound with $\mathcal{W}=2$.}
\label{fig:downWn0}
\end {figure}

\begin {figure}[!t]
\centering
\includegraphics[width=\linewidth]{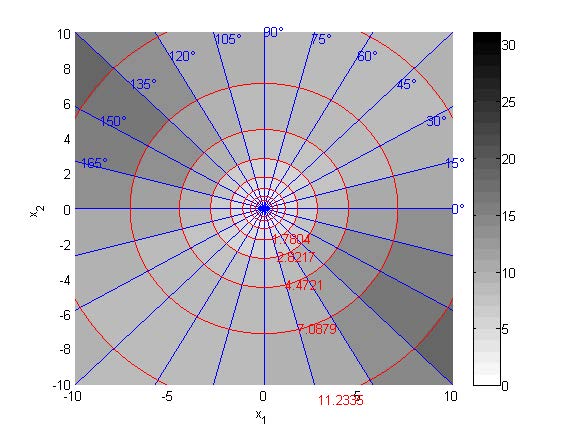}
\caption{Zoomed-in result of the regional lower bound with $\mathcal{W}=2$.}
\label{fig:downWn0zoom}
\end {figure}

\begin {figure}[!t]
\centering
\includegraphics[width=\linewidth]{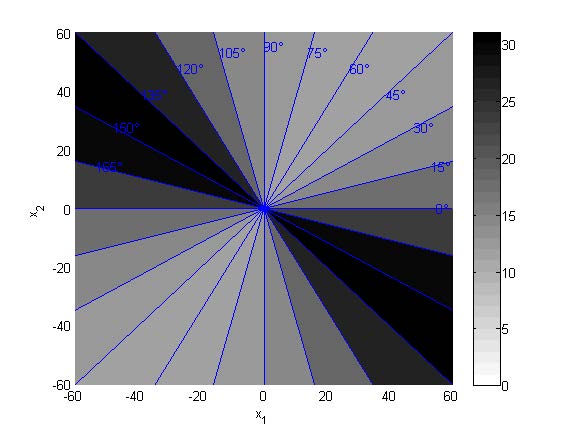}
\caption{Computed result of the regional upper bound with $w=0$.}
\label{fig:up}
\end {figure}

We set $\mathcal{W}=2$, the convergence rate $\rho=0.01$, $\mathcal{L}_2$ gain $\gamma=2$, sampling time $h=0.005s$, event condition $\sigma=0.1$. By checking the LMI presented in \cite{heemels2013periodic}, we can see there exists a feasible solution, thus the stability and performance can be guaranteed. The result of the computed lower bound by Theorem \ref{theorem:down} is shown in Figure \ref{fig:downWn0}. Figure \ref{fig:downWn0zoom} shows a zoomed-in version. The computed upper bound by Corollary \ref{theorem:up} is shown in Figure \ref{fig:up}. The resulting abstraction precision is $\epsilon=0.15s$.

\begin {figure}[!t]
\centering
\includegraphics[width=\linewidth]{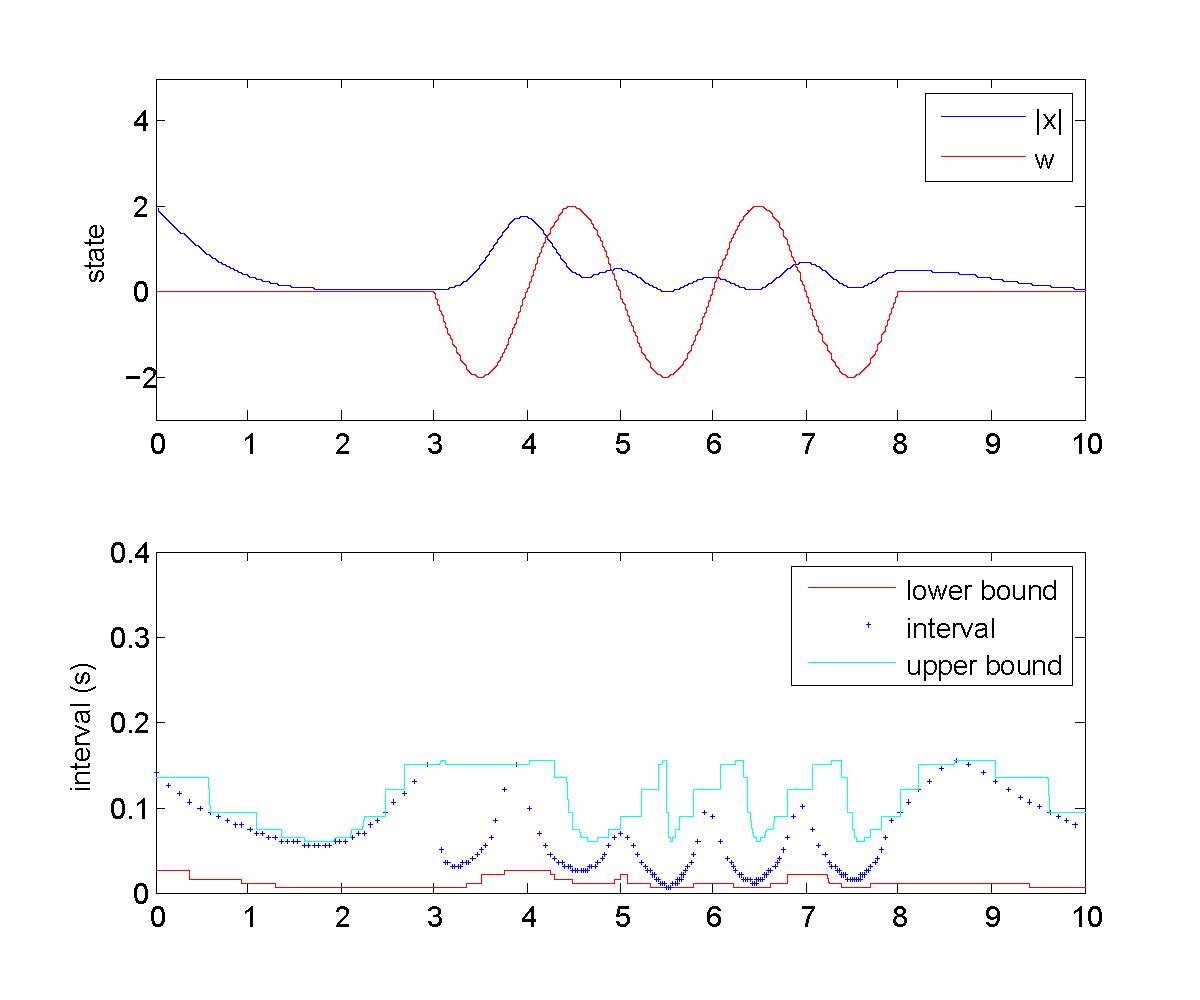}
\caption{System evolution and event intervals when $w=2\sin(\pi t),t\in[3,8]$: state evaluation and perturbance, event intervals with the bounds.}
\label{fig:evoWn0}
\end {figure}

The simulation results of the system evolution and event intervals with perturbations are shown in Figure \ref{fig:evoWn0}. The upper bound triggered 6 events during the $10s$ simulation. Note that, increasing the number of subdivisions can lead to less conserved lower and upper bounds of the inter event time. The conservativeness can also be reduced by decreasing $\mathcal{W}$.

\begin {figure}[!t]
\centering
\includegraphics[width=\linewidth]{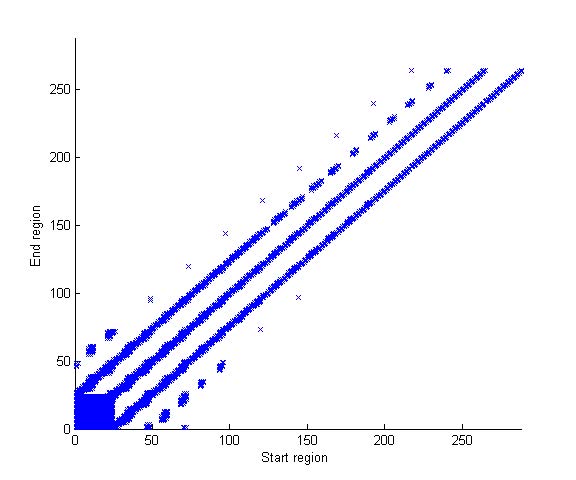}
\caption{Reachable regions starting from each state region, with labeling from Figure \ref{fig:gridpartition}.}
\label{fig:gridflowpipe}
\end {figure}

\begin {figure}[!t]
\centering
\includegraphics[width=\linewidth]{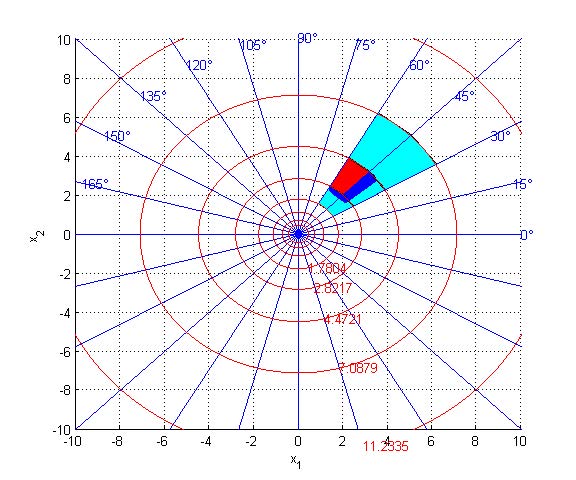}
\caption{Flow pipe of $(s_1,s_2)=(4,6)$: indicating initial state set (red), reachable state set (blue), and reachable regions (cyan).}
\label{fig:flowpipe46}
\end {figure}

The reachable state regions starting from each region is shown in Figure \ref{fig:gridflowpipe}. As an example, the reachable state region of the initial region $(s_1,s_2)=(4,6)$ is shown in Figure \ref{fig:flowpipe46}.

\begin {figure}[!t]
\centering
\includegraphics[width=\linewidth]{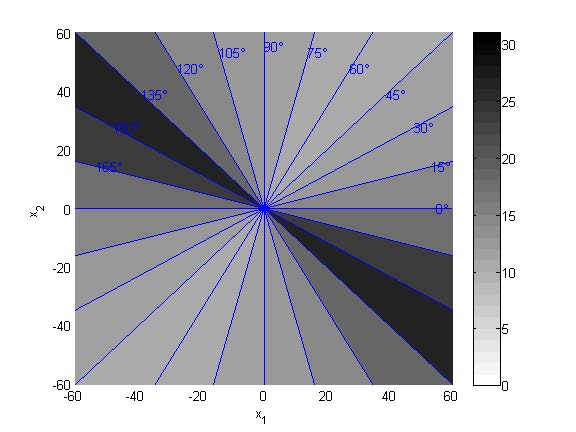}
\caption{Computed result of the regional lower bound with $w=0$.}
\label{fig:lowerboundW0}
\end {figure}

\begin {figure}[!t]
\centering
\includegraphics[width=\linewidth]{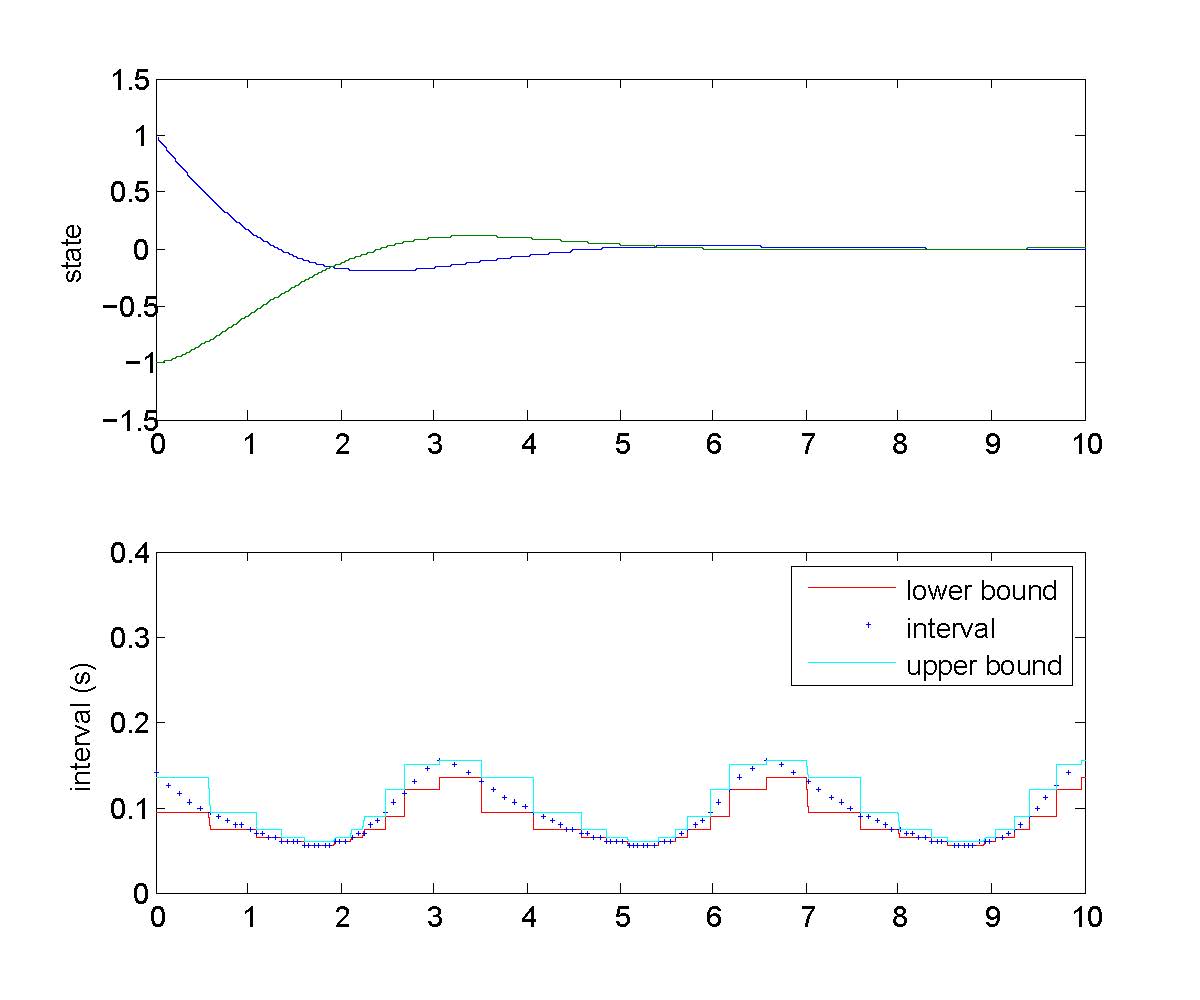}
\caption{System evolution and event intervals when $w=0$: state evaluation and event intervals vs computed bounds.}
\label{fig:evoW0}
\end {figure}

\begin {figure}[!t]
\centering
\includegraphics[width=\linewidth]{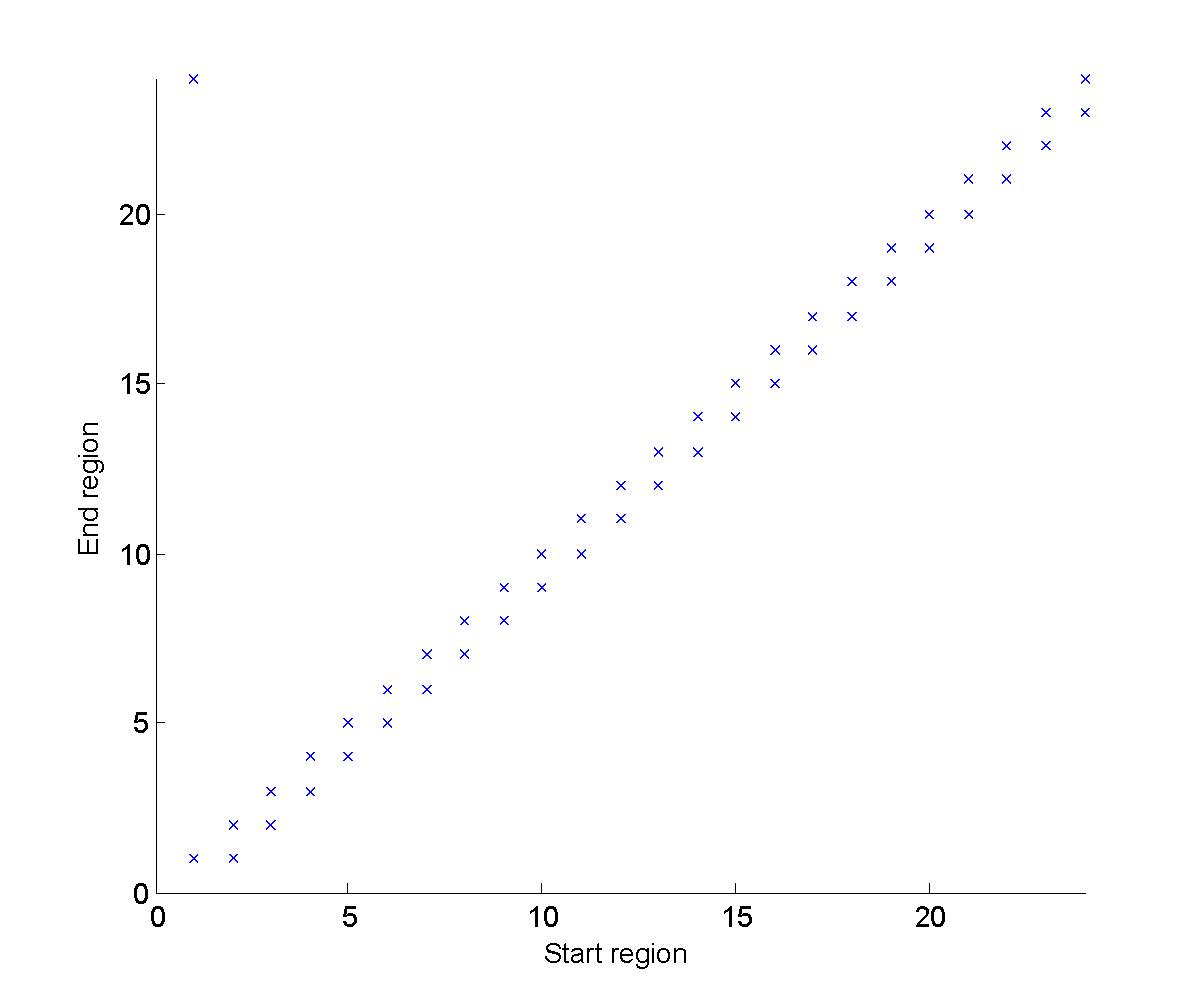}
\caption{Reachable regions starting from each conic region, with labeling from Figure \ref{fig:gridpartition}.}
\label{fig:coneflowpipe}
\end {figure}

We also present a simulation when $w=0$. The lower bound is shown in Figure \ref{fig:lowerboundW0}. The evolution of the system is shown in Figure \ref{fig:evoW0}, which shows that, the inter event intervals are within the computed bounds. The reachable state regions starting from each region are shown in Figure \ref{fig:coneflowpipe}.

\section{Conclusion}
\label{section:conclusion}

In this paper, we present a construction of a power quotient system for the traffic model of the PETC implementations from \cite{heemels2013periodic}. The constructed models can be used to estimate the next event time and the state set when the next event occurs. These models allow to design scheduling to improve listening time of wireless communications and medium access time to increase the energy consumption and bandwidth occupation efficiency.

In this paper, we consider an output feedback system with a dynamic controller. However, the state partition is still based on the states of the system and controller. The system state may not always be obtainable. Therefore, to estimate the system state in an ETC implementation from output measurements is a very important extension to make this work more practical. The periodic asynchronous event-triggered control (PAETC) presented in \cite{FuM17PAETC} is an extension of PETC considering quantization. One can either treat the quantization error as part of the perturbations, or analyze this part separately to increase the abstraction precision, since the dynamics of the quantization error is dependent on the states. This is also an interesting future investigation. Another interesting extension is reconstruction of traffic models for each sensor node to capture the local timing behavior in a decentralized PETC implementation, by either global information or even only local information.

\section*{Appendix}

\begin{IEEEproof}[\textbf{Isotropic covering}]
Consider $x=\begin{bmatrix}
              x_1 & x_2 & \cdots & x_n \\
            \end{bmatrix}^{\mathrm{T}}
\in\mathbb{R}^{n}$. We first present a case when $x\in\mathbb{R}^2$. Let $\Theta=[-\frac{\pi}{2},\frac{\pi}{2}[$ be an interval. Splitting this interval into $q$ sub-intervals and $\Theta_s=[\underline{\theta}_s,\overline{\theta}_s[$ be the $s$-th sub-interval. Then for each sub-interval, one can construct a cone pointing at the origin:
\begin{equation*}
\mathcal{R}_s=\left\{x\in\mathbb{R}^2|x^{\mathrm{T}}\tilde\Xi_s x\geq 0\right\},
\end{equation*}
where
\begin{equation*}
\tilde\Xi_s=\begin{bmatrix}
      -\sin\underline{\theta}_s\sin\overline\theta_s & \frac{1}{2}\sin(\underline\theta_s+\overline\theta_s) \\
      \frac{1}{2}\sin(\underline\theta_s+\overline\theta_s) & -\cos\underline\theta_s\cos\overline\theta_s \\
    \end{bmatrix}.
\end{equation*}
Remark \ref{remark:weq0} shows that $x$ and $-x$ have the same behaviours, therefore it is sufficient to only consider half of the state-space.

Now we derive the case when $x\in\mathbb{R}^n$, $n>2$. Define $(x)_{i,j}=(x_i,x_j)$ as the projection of this point on its $i-j$ coordinate. Now a polyhedral cone $\mathcal{R}_{s}$ can be defined as:
\begin{equation*}
\mathcal{R}_{s}=\left\{x\in\mathbb{R}^n\left| \bigwedge_{i=1}^{n-1}(x)^{\mathrm{T}}_{(i,i+1)}\tilde{\Xi}_{s,(i,i+1)}(x)_{(i,i+1)}\geq 0\right.\right\},
\end{equation*}
where $\tilde{\Xi}_{s,(i,i+1)}$ is a constructed matrix. A relation between $\tilde{\Xi}_{s,(i,i+1)}$ and $\Xi_{s,(i,i+1)}$ from (\ref{eq:stateset}) is given by:
\begin{equation*}
\left\{
\begin{aligned}
&[\Xi_{s,(i,i+1)}]_{(i,i)}&=&[\tilde{\Xi}_{s,(i,i+1)}]_{(1,1)}\\
&[\Xi_{s,(i,i+1)}]_{(i,i+1)}&=&[\tilde{\Xi}_{s,(i,i+1)}]_{(1,2)}\\
&[\Xi_{s,(i,i+1)}]_{(i+1,i)}&=&[\tilde{\Xi}_{s,(i,i+1)}]_{(2,1)}\\
&[\Xi_{s,(i,i+1)}]_{(i+1,i+1)}&=&[\tilde{\Xi}_{s,(i,i+1)}]_{(2,2)}\\
&[\Xi_{s,(i,i+1)}]_{(k,l)}&=&0,
\end{aligned}\right.
\end{equation*}
where $[M]_{(i,j)}$ is the $i$-th row, $j$-th column entry of the matrix $M$, $k$ and $l$ satisfy $(k,l)\neq (i,i+1)$.
\end{IEEEproof}

\begin{IEEEproof}[\textbf{Proof of Lemma \ref{lemma:phiapproximate}}]
We decouple the event triggering mechanism in (\ref{eq:underlinekx}) first:
\begin{equation}\label{eq:phi1}
\begin{aligned}
&\begin{bmatrix}
M(k)x+\Theta(k) \\
C_Ex \\
\end{bmatrix}^{\mathrm{T}}Q
\begin{bmatrix}
M(k)x+\Theta(k) \\
C_Ex \\
\end{bmatrix}\\
=&x^{\mathrm{T}}\Phi_1(k)x+x^{\mathrm{T}}\Phi_2(k)\Theta(k)+ \Theta^{\mathrm{T}}(k)\Phi_2^{\mathrm{T}}(k)x\\
&+\Theta^{\mathrm{T}}(k)Q_1\Theta(k)\\
\leq&x^{\mathrm{T}}(\Phi_1(k)+\Phi_2(k)\Psi^{-1}\Phi_2^{\mathrm{T}}(k))x+ \Theta^{\mathrm{T}}(k)(Q_1+\Psi)\Theta(k),
\end{aligned}
\end{equation}
where the last inequality comes from Lemma 6.2 in \cite{gu2003stability}. Now for the uncertainty part, we have:
\begin{equation*}
\begin{aligned}
&\Theta^{\mathrm{T}}(k)(Q_1+\Psi)\Theta(k)\\
=&\begin{bmatrix}
        \Theta_1(k) \\
        \mathbf{0} \\
      \end{bmatrix}^{\mathrm{T}}
      \begin{bmatrix}
        (Q_1+\Psi)_1 & (Q_1+\Psi)_2 \\
        (Q_1+\Psi)_3 & (Q_1+\Psi)_4 \\
      \end{bmatrix}
      \begin{bmatrix}
        \Theta_1(k) \\
        \mathbf{0} \\
      \end{bmatrix}\\
=&\Theta_1^{\mathrm{T}}(k)(Q_1+\Psi)_1\Theta_1(k).
\end{aligned}
\end{equation*}
From the hypothesis of the theorem that there exists $\mu$ such that $(Q_1+\Psi)_1\preceq\mu I$, together with Jensen's inequality \cite{gu2003stability}, inequality (2.2) in \cite{van1977sensitivity}, and Assumption \ref{assumption:w}, i.e. $w\in\mathcal{L}_{\infty}$, $\Theta^{\mathrm{T}}(k)(Q_1+\Psi)\Theta(k)$ can be bounded from above by:
\begin{equation}\label{eq:phi3}
\begin{aligned}
&\Theta^{\mathrm{T}}(k)(Q_1+\Psi)\Theta(k)=\Theta_1^{\mathrm{T}}(k)(Q_1+\Psi)_1\Theta_1(k)\\
\leq&\mu\left(\int_0^{kh}e^{A_p(kh-s)}Ew(s)ds\right)^{\mathrm{T}}\left(\int_0^{kh}e^{A_p(kh-s)}Ew(s)ds\right)\,\\
&\left(\text{by } (Q_1+\Psi)\preceq\mu I\right)\\
\leq& kh\mu\int_0^{kh}\left(e^{A_p(kh-s)}Ew(s)\right)^{\mathrm{T}}\left(e^{A_p(kh-s)}Ew(s)\right)ds\,\\
&(\text{by Jensen's equality})\\
\leq& kh\mu\int_0^{kh}e^{(kh-s)\lambda_{\max}\left(A_p+A_p^{\mathrm{T}}\right)}w^{\mathrm{T}}(s) E^{\mathrm{T}}Ew(s)ds\,\\
&(\text{by (2.2) in \cite{van1977sensitivity}})\\
\leq& kh\mu\lambda_{\max}\left(E^{\mathrm{T}}E\right)\int_0^{kh}e^{(kh-s)\lambda_{\max}\left(A_p+ A_p^{\mathrm{T}}\right)}ds\|w\|^2_{\mathcal{L}_{\infty}}\,\\
&(\text{by }w\in\mathcal{L}_{\infty})\\
=&kh\mu\lambda_{\max}\left(E^{\mathrm{T}}E\right)d_{A_p}(k)\|w\|^2_{\mathcal{L}_{\infty}}.
\end{aligned}
\end{equation}
With (\ref{eq:phi3}), (\ref{eq:phi1}) can be further bounded as:
\begin{equation}\label{eq:phi4}
\begin{aligned}
&\begin{bmatrix}
M(k)x+\Theta(k) \\
C_Ex \\
\end{bmatrix}^{\mathrm{T}}Q
\begin{bmatrix}
M(k)x+\Theta(k) \\
C_Ex \\
\end{bmatrix}\\
\leq&x^{\mathrm{T}}\left(\Phi_1(k)+\Phi_2(k)\Psi^{-1}\Phi_2^{\mathrm{T}}(k)\right)x+\Phi_3(k)\|w\|^2_{\mathcal{L}_{\infty}}.
\end{aligned}
\end{equation}
From the hypothesis of the theorem, if $\Phi(k)\preceq 0$ holds, then by applying the Schur complement to (\ref{eq:Ephi}), the following inequality holds:
\begin{equation*}
x^{\mathrm{T}}\left(\Phi_1(k)+\Phi_2(k)\Psi^{-1}\Phi_2^{\mathrm{T}}(k)\right)x+\Phi_3(k)\|w\|^2_{\mathcal{L}_{\infty}}\leq 0,
\end{equation*}
which indicates:
\begin{equation}\label{eq:phi5}
\begin{aligned}
\begin{bmatrix}
M(k)x+\Theta(k) \\
C_Ex \\
\end{bmatrix}^{\mathrm{T}}Q
\begin{bmatrix}
M(k)x+\Theta(k) \\
C_Ex \\
\end{bmatrix}\leq 0.
\end{aligned}
\end{equation}
Therefore, $\underline{k}_x$ generated by (\ref{eq:underlinekx}) is lower bounded by $k_x'$ generated by (\ref{eq:phiapproximate}). This ends the proof.
\end{IEEEproof}

\begin{IEEEproof}[\textbf{Proof of Theorem \ref{theorem:down}}]
We first consider the regions with $s_2>1$. If all the hypothesis of the theorem hold, by applying the Schur complement to (\ref{eq:mainLMI}), one has:
\begin{equation}\label{eq:theorem1}
\begin{aligned}
x^{\mathrm{T}}\left(H+\Phi_2(k)\Psi^{-1}\Phi_2^{\mathrm{T}}(k) \right)x\leq0.
\end{aligned}
\end{equation}
From (\ref{eq:stateset}), and applying the S-procedure, it holds that:
\begin{equation}\label{eq:theorem2}
\begin{aligned}
x^{\mathrm{T}}\left(\Phi_1(k)+\Phi_3(k)\mathcal{W}^2W_{s_2-1}^{-2}I+\Phi_2(k) \Psi^{-1}\Phi_2^{\mathrm{T}}(k)\right)x\leq0.
\end{aligned}
\end{equation}
From (\ref{eq:stateset}) we also have:
\begin{equation}\label{eq:ph}
x^{\mathrm{T}}x\geq W_{s_2-1}^2.
\end{equation}
Since $\Phi_3(k)$, $\mathcal{W}$, and $W_{s_2-1}$ are non-negative scalars and $W_{s_2-1}>0$, we have the following inequality:
\begin{equation}\label{eq:xw}
\begin{aligned}
&x^{\mathrm{T}}\Phi_3(k)\mathcal{W}^2W_{s_2-1}^{-2}Ix=\Phi_3(k)\mathcal{W}^2W_{s_2-1}^{-2}x^{\mathrm{T}}x\\
\geq&\Phi_3(k)\mathcal{W}^2W_{s_2-1}^{-2}W_{s_2-1}^{2}=\Phi_3(k)\mathcal{W}^2\geq\Phi_3(k)\|w\|^2_{\mathcal{L}_{\infty}},
\end{aligned}
\end{equation}
in which the last inequality comes form the definition of $\mathcal{W}$. Now inserting (\ref{eq:xw}) into (\ref{eq:theorem2}) results in:
\begin{equation*}
x^{\mathrm{T}}\left(\Phi_1(k)+\Phi_2(k)\Psi^{-1}\Phi_2^{\mathrm{T}}(k)\right)x+\Phi_3(k)\|w\|^2_{\mathcal{L}_{\infty}}\leq 0,
\end{equation*}
which together with applying the Schur complement to (\ref{eq:Ephi}) provides the regional lower bound.

When $s_2=1$, $k>0$, $H$ will be diagonal infinity. Thus the LMI (\ref{eq:mainLMI}) will be infeasible. According to the event-triggered condition (\ref{eq:newcondition}), which indicates that $t_{b+1}\in\mathcal{T}_s$ and $t_{b+1}>t_{b}$, the regional lower bound for those regions with $s_2=1$ is $h$. This finishes the proof.
\end{IEEEproof}

\begin{IEEEproof}[\textbf{Proof of Corollary \ref{corollary:downw0}}]
The result can be easily obtained from Theorem \ref{theorem:down} considering $E=\mathbf{0}$.
\end{IEEEproof}

\begin{IEEEproof}[\textbf{Proof of Corollary \ref{theorem:up}}]
The result can be easily obtained analogously to Theorem \ref{theorem:down} considering $E=\mathbf{0}$: if all the hypothesis of this Corollary hold, then according to (\ref{eq:upperLMI}), $\Phi_1(k)\succ 0$, $k\in\left\{\bar{k}_{s_1,\bullet},\cdots,\bar{l}\right\}$. According to the definition of $\Phi_1(k)$ in (\ref{eq:Phi123}), for all $k\geq \bar{k}_{s_1,\bullet}$, it holds that:
\begin{equation*}
\begin{bmatrix}
                                     M(k)x \\
                                     C_Ex \\
                                   \end{bmatrix}^{\mathrm{T}}Q
                                   \begin{bmatrix}
                                     M(k)x \\
                                     C_Ex \\
                                   \end{bmatrix}>0,
\end{equation*}
which together with event condition (\ref{eq:nonxeventsequence}) provides the regional upper bound.
\end{IEEEproof}

\begin{IEEEproof}[\textbf{Proof of Theorem \ref{theorem:mainresult}}]
The result follows from Lemma \ref{lemma:approximatelysimulation} and the construction described in this section.
\end{IEEEproof}

\ifCLASSOPTIONcaptionsoff
  \newpage
\fi

\bibliographystyle{plain}        
\bibliography{mybib}           

\begin{thebibliography}{10}

\bibitem{chutinan1998computing}
Alongkrit Chutinan and Bruce~H Krogh.
\newblock Computing polyhedral approximations to flow pipes for dynamic
  systems.
\newblock In {\em Decision and Control, 1998. Proceedings of the 37th IEEE
  Conference on}, volume~2, pages 2089--2094. IEEE, 1998.

\bibitem{chutinan2003computational}
Alongkrit Chutinan and Bruce~H Krogh.
\newblock Computational techniques for hybrid system verification.
\newblock {\em IEEE transactions on automatic control}, 48(1):64--75, 2003.

\bibitem{cloosterman2010controller}
Marieke~BG Cloosterman, Laurentiu Hetel, Nathan Van~de Wouw, WPMH Heemels,
  Jamal Daafouz, and Henk Nijmeijer.
\newblock Controller synthesis for networked control systems.
\newblock {\em Automatica}, 46(10):1584--1594, 2010.

\bibitem{cloosterman2009stability}
Marieke~BG Cloosterman, Nathan Van~de Wouw, WPMH Heemels, and Hendrik
  Nijmeijer.
\newblock Stability of networked control systems with uncertain time-varying
  delays.
\newblock {\em IEEE Transactions on Automatic Control}, 54(7):1575--1580, 2009.

\bibitem{donkers2012output}
MCF Donkers and WPMH Heemels.
\newblock Output-based event-triggered control with guaranteed-gain and
  improved and decentralized event-triggering.
\newblock {\em Automatic Control, IEEE Transactions on}, 57(6):1362--1376,
  2012.

\bibitem{donkers2011stability}
MCF Donkers, WPMH Heemels, Nathan Van~de Wouw, and Laurentiu Hetel.
\newblock Stability analysis of networked control systems using a switched
  linear systems approach.
\newblock {\em IEEE Transactions on Automatic control}, 56(9):2101--2115, 2011.

\bibitem{ewald2012combinatorial}
G{\"u}nter Ewald.
\newblock {\em Combinatorial convexity and algebraic geometry}, volume 168.
\newblock Springer Science \& Business Media, 2012.

\bibitem{fiter2012state}
Christophe Fiter, Laurentiu Hetel, Wilfrid Perruquetti, and Jean-Pierre
  Richard.
\newblock A state dependent sampling for linear state feedback.
\newblock {\em Automatica}, 48(8):1860--1867, 2012.

\bibitem{fiter2015robust}
Christophe Fiter, Laurentiu Hetel, Wilfrid Perruquetti, and Jean-Pierre
  Richard.
\newblock A robust stability framework for lti systems with time-varying
  sampling.
\newblock {\em Automatica}, 54:56--64, 2015.

\bibitem{FuM17PAETC}
Anqi Fu and Manuel~Mazo Jr.
\newblock Periodic asynchronous event-triggered control.
\newblock {\em CoRR}, abs/1703.10073, 2017.

\bibitem{gielen2010polytopic}
Rob~H Gielen, Sorin Olaru, Mircea Lazar, WPMH Heemels, Nathan van~de Wouw, and
  S-I Niculescu.
\newblock On polytopic inclusions as a modeling framework for systems with
  time-varying delays.
\newblock {\em Automatica}, 46(3):615--619, 2010.

\bibitem{gu2003stability}
Keqin Gu, Jie Chen, and Vladimir~L Kharitonov.
\newblock {\em Stability of time-delay systems}.
\newblock Springer Science \& Business Media, 2003.

\bibitem{heemels2013periodic}
WPMH Heemels, MCF Donkers, and Andrew~R Teel.
\newblock Periodic event-triggered control for linear systems.
\newblock {\em Automatic Control, IEEE Transactions on}, 58(4):847--861, 2013.

\bibitem{hetel2006stabilization}
Laurentiu Hetel, Jamal Daafouz, and Claude Iung.
\newblock Stabilization of arbitrary switched linear systems with unknown
  time-varying delays.
\newblock {\em IEEE Transactions on Automatic Control}, 51(10):1668--1674,
  2006.

\bibitem{kolarijani2015symbolic}
Arman~Sharifi Kolarijani, Dieky Adzkiya, and Manuel Mazo.
\newblock Symbolic abstractions for the scheduling of event-triggered control
  systems.
\newblock In {\em Decision and Control (CDC), 2015 IEEE 54th Annual Conference
  on}, pages 6153--6158. IEEE, 2015.

\bibitem{kolarijani2016formal}
Arman~Sharifi Kolarijani and Manuel Mazo~Jr.
\newblock A formal traffic characterization of lti event-triggered control
  systems.
\newblock {\em IEEE Transactions on Control of Network Systems}, 2016.

\bibitem{MazoJr2014}
Manuel Mazo~Jr. and Ming Cao.
\newblock Asynchronous decentralized event-triggered control.
\newblock {\em Automatica}, 50(12):3197--3203, 2014.

\bibitem{mazo2015decentralized}
Manuel Mazo~Jr and Anqi Fu.
\newblock Decentralized event-triggered controller implementations.
\newblock {\em Event-Based Control and Signal Processing}, page 121, 2015.

\bibitem{mazo2011decentralizedtac}
Manuel Mazo~Jr. and Paulo Tabuada.
\newblock Decentralized event-triggered control over wireless sensor/actuator
  networks.
\newblock {\em Automatic Control, IEEE Transactions on}, 56(10):2456--2461,
  2011.

\bibitem{skaf2009analysis}
Jo{\"e}lle Skaf and Stephen Boyd.
\newblock Analysis and synthesis of state-feedback controllers with timing
  jitter.
\newblock {\em IEEE Transactions on Automatic Control}, 54(3):652--657, 2009.

\bibitem{suh2008stability}
Young~Soo Suh.
\newblock Stability and stabilization of nonuniform sampling systems.
\newblock {\em Automatica}, 44(12):3222--3226, 2008.

\bibitem{tabuada2007event}
Paulo Tabuada.
\newblock Event-triggered real-time scheduling of stabilizing control tasks.
\newblock {\em Automatic Control, IEEE Transactions on}, 52(9):1680--1685,
  2007.

\bibitem{tabuada2009verification}
Paulo Tabuada.
\newblock {\em Verification and control of hybrid systems: a symbolic
  approach}.
\newblock Springer Science \& Business Media, 2009.

\bibitem{van1977sensitivity}
Charles Van~Loan.
\newblock The sensitivity of the matrix exponential.
\newblock {\em SIAM Journal on Numerical Analysis}, 14(6):971--981, 1977.

\bibitem{wang2011event}
Xiaofeng Wang and Michael~D Lemmon.
\newblock Event-triggering in distributed networked control systems.
\newblock {\em Automatic Control, IEEE Transactions on}, 56(3):586--601, 2011.

\bibitem{wang2011eventautomatica}
Xiaofeng Wang and Michael~D Lemmon.
\newblock On event design in event-triggered feedback systems.
\newblock {\em Automatica}, 47(10):2319--2322, 2011.

\end{thebibliography}

\end{document}